\documentclass[useAMS,usenatbib]{mn2e}
\usepackage{graphicx}
\usepackage{amsmath}
\usepackage{array}

\title[The Size of Star Clusters Accreted by the Milky Way]{The Size of Star Clusters Accreted by the Milky Way}
\author[Meghan Miholics, Jeremy J. Webb, Alison Sills]{Meghan Miholics\thanks{E-mail: miholim@mcmaster.ca (MM)}, Jeremy J. Webb, Alison Sills\\
Department of Physics and Astronomy, McMaster University, Hamilton ON, L8S 4M1,Canada}
\begin{document}


\pagerange{\pageref{firstpage}--\pageref{lastpage}} \pubyear{2014}

\maketitle

\label{firstpage}

\begin{abstract}
We perform $N$-body simulations of a cluster that forms in a dwarf galaxy and is then accreted by the Milky Way to investigate how a cluster's structure is affected by a galaxy merger. We find that the cluster's half mass radius will respond quickly to this change in potential. When the cluster is placed on an orbit in the Milky Way with a stronger tidal field the cluster experiences a sharp decrease in size in response to increased tidal forces. Conversely, when placed on an orbit with a weaker tidal field the cluster expands since tidal forces decrease and stars moving outwards due to internal effects remain bound at further distances than before. In all cases, we find that the cluster's half mass radius will eventually be indistinguishable from a cluster that has always lived in the Milky Way on that orbit. These adjustments occur within 1-2 half mass relaxation times of the cluster in the dwarf galaxy. We also find this effect to be qualitatively independent of the time that the cluster is taken from the dwarf galaxy. In contrast to the half mass radius, we show the core radius of the cluster is not affected by the potential the cluster lives in. Our work suggests that structural properties of accreted clusters are not distinct from clusters born in the Milky Way. Other cluster properties, such as metallicity and horizontal branch morphology, may be the only way to identify accreted star clusters in the Milky Way.
\end{abstract}


\section{Introduction}

	All galaxies are thought to grow in a hierarchical fashion in which galaxies begin small and merge to form larger ones \citep{b29}. Globular clusters are tracers of these processes since they are some of the oldest structures in our universe and are found in almost all types of galaxies. The Milky Way itself has not experienced any major mergers ($>$1:10) since its formation \citep{Ruchti14} but has been built up from mergers with smaller dwarf galaxies \citep{b24}. Currently the Milky Way has a number of dwarf galaxy satellites which could merge with it in the future. Globular clusters are known to exist in at least three of these systems: LMC, SMC, and Fornax \citep{b6}. If a dwarf galaxy merges with the Milky Way its globular clusters are expected to be incorporated into the Milky Way's globular cluster population. Direct evidence for this scenario is provided by the Sagittarius dwarf spheriodal galaxy that is in the process of merging with the Milky Way \citep{b11}. Using distance and velocity measurements of stars in Sagittarius and of globular clusters in the Milky Way, \citet{b14} determined that 5 of the Milky Way's globular clusters likely originated in Sagittarius. Additional evidence for the accretion of globular clusters through mergers has been provided through comparison of the Milky Way globular clusters with globular clusters in satellite galaxies. \citet{b32} separated the Milky Way's halo globular clusters into two groups, the Young and Old Halo populations, based on horizontal branch (HB) morphology. \citet{b32} showed that these populations have distinct radial distributions and kinematics from one another and hypothesized that the Young Halo population originated from dwarf galaxies that merged with the Milky Way. \citet{b15} extended this work by comparing the HB morphologies of the Young and Old Halo populations to the clusters in the four aforementioned satellite galaxies. They found that clusters in the Milky Way's satellites have HB morphologies similar to the Young Halo population, excluding approximately 20-30\% which are similar to the Old Halo population. \citet{b15} also compared the distribution of cluster core radii for the Young Halo, Old Halo and external populations. They found that while the Old Halo and Young Halo clusters  have dissimilar distributions in core radii the distribution of Young Halo core radii and the core radii of clusters of external origin are quite similar. They concluded that all 30 of the Young Halo clusters as well as 10-12 Old Halo clusters originated in dwarf galaxies and estimated that the Milky Way has merged with approximately 7 dwarf spheroidal type galaxies. Other evidence for accretion of clusters from dwarf galaxies includes the age-metallicity relation (AMR) of the globular clusters in the Milky Way. \citet{b5} demonstrated that two different tracks in the AMR of Milky Way clusters exist, one at a constant age of approximately 12.8 Gyr over a whole range of metallicities and another with younger ages for a given metallicity. They showed that the AMR of the globular clusters commonly associated with Sagittarius and the Canis Major systems (another potentially merging dwarf galaxy \citet{b17}) is similar to the latter. Combining this finding with information about the motions of some globular clusters in the Milky Way, they concluded that 27-47 of the Milky Way's globular clusters have been accreted, corresponding to mergers with 6-8 dwarf galaxies. \\

	The size of a cluster is sometimes used as evidence that certain clusters originated in a dwarf galaxy. As previously mentioned, \citet{b15} directly compared the core radii of Milky Way clusters with those of satellite galaxy clusters. Additionally, a large half mass radius is sometimes assumed to be evidence that a cluster may have been accreted from a dwarf galaxy, as was recently suggested for Palomar 14 \citep{Frank}. However, it is not clear how the core radius, half mass radius, or any alternative measure of size would change during a galaxy merger. The size of a cluster in an external potential should be regulated by tidal effects. An observational measure for the size of a cluster is the limiting radius, which is the radius where the surface brightness profile goes to zero \citep{b12}. Another measure of cluster size often used is the Jacobi radius, $r_J$, which is given by the distance between the cluster's centre and the first Lagrange point of cluster in the galaxy's potential \citep{b4}. Roughly speaking, it is the radius at which a cluster star feels an equal force from the cluster and the galaxy in which it orbits \citep{b27}. The Jacobi radius is usually larger than the limiting radius since stars with positive total energy outside of the Jacobi radius are stripped away from the cluster. For a spherically symmetric potential $r_J$ is given, to first order, by:
\begin{equation}
\label{eq:rj}
r_J = R_{gc} (\frac{M_c}{2M_g})^{1/3}
\end{equation}
where $R_{gc}$ is the galactocentric radius of the cluster, $M_c$ is the mass of the cluster and $M_g$ is the mass of the galaxy enclosed in the cluster's orbit \citep{b27}. \\

	During a galaxy merger, the cluster undergoes a change in potential which could cause a change in the cluster's size. However, it is not clear over what timescale these changes will occur and what their magnitude will be. Although changes in size have not been explicitly studied in detail, some work has been done to look at mass loss of clusters in galaxy mergers. Most recently, \citet{b16} performed simulations of clusters in disk galaxies with different masses. They placed several of these simulations together, each time increasing the galaxy's mass, to represent a galaxy that is undergoing ``instantaneous" mergers. They found that each merger event increases the mass loss rate of the cluster and that dissolution times are shorter for clusters in a disk with a constant mass than clusters that evolve in a galaxy that builds up the same mass through mergers. Efforts have also been made to simulate clusters in a galaxy merger using a continuously changing potential. \citet{b23} tracked the tidal history of clusters in a cosmological context by inserting clusters into a dark matter only $\Lambda$CDM simulation. They also found that mergers tend to increase the mass loss rates of clusters. Similar work was done by \citet{b21} who conducted simulations of clusters on different orbits in a major merger of galaxies emulating NGC 4038/39. These simulations were done using a new version of \textsc{nbody6} \citep{b1,Aarseth03} that calculates the tidal force along the cluster's orbit in the potential \citep{b22}. They found that mass loss rates depend strongly on the orbit of the cluster; clusters that remain bound to the merger remnant experience increases in mass loss rates in comparison to evolution in an isolated galaxy, while clusters that are ejected into tidal debris have greatly reduced mass loss rates. They also noted that the half mass radius of a cluster that stays bound to the remnant starts to decrease after the final coalescence and the half mass radius of a cluster in the tidal debris expands as though the cluster was in isolation.\\

	In this work, we will examine how the size of a dwarf galaxy cluster would change if it is accreted by the Milky Way in a merger event to determine if the size of a cluster can be used to tell where it came from. We perform $N$-body simulations of a cluster that starts in a dwarf galaxy potential and is then instantaneously switched to a Milky Way potential to emulate a galaxy merger event. We use the cluster's half mass radius, $r_h$, as our measure of size since it traces the commonly observed half light radius. We examine the effect of moving the cluster at different ages, as well as the effect of giving the cluster different galactocentric distances in the Milky Way. \\

\section{Methods}

We use \textsc{nbody6} \citep{b1,Aarseth03} to carry out collisional simulations of a star cluster that starts in a dwarf galaxy potential and is switched to the Milky Way's potential. The code calculates the gravitational force between all stars in the cluster as well as the force between the cluster and the galaxy, which is modeled as a static potential. The effects of stellar evolution are also implemented throughout the simulation. All clusters initially have 50,000 stars with zero binary stars and an average mass per star of 0.6 $M_{\odot}$. Initial masses for the stars are assigned to follow a Kroupa initial mass function \citep{b13}. Initial positions of the stars in the cluster are such that the mass density as a function of distance is given by a Plummer profile \citep{b19}. The parameter that is used to set the size of the cluster for the simulation is the virial radius which we assign an initial value of 4.0 pc. This initial virial radius corresponds to a half mass radius of 3.2 pc \citep{b10}. Initial velocities are assigned such that the cluster is in virial equilibrium.\\

The second component of our simulations is the model for the potential of the galaxy. For the current simulations we take advantage of the code's ability to model an external static disk galaxy (such as the Milky Way) using three components: a central bulge, disk and dark matter halo. The bulge is modeled as a point mass at the centre of the galaxy while the disk has a gravitational potential according to \cite{b18}:
\begin{equation}
\Phi_D(R,z) = -\frac{GM_{disk}}{(R^2 + [a+(z^2+b^2)^{1/2}]^2)^{1/2}}
\end{equation}
where $M_{disk}$ is the total mass of the disk, $R$ and $z$ are the cylindrical coordinates describing distance from the centre of the disk and height above the disk, and $a$ and $b$ are parameters that adjust the shape of the disk. The halo is represented by a logarithmic potential given by:
\begin{equation}
\Phi_H = \frac{1}{2} V_{CIRC}^2\ln(r^2 + r_o^2)
\end{equation}
where r is galactocentric radius. $V_{CIRC}$ is the velocity of a body on a circular orbit of radius $R_{CIRC}$ in the total potential provided by the bulge, disk and halo. $r_o$ is the radius at which an object on a circular orbit would have velocity $V_{CIRC}$ if only the halo potential is considered. $V_{CIRC}$ and $R_{CIRC}$ are chosen to emulate the galaxy that is being modelled while $r_o$ is calculated based on the contributions from the bulge and disk.\\

\begin{table}
\caption{Parameters used to model each galaxy. References: (1) \citet{b30}; (2) \citet{b20}; (3) \citet{b31}; (4) \citet{b26}; (5) This work (6) Calculated. }
\label{tab:parameters}
\begin{tabular}{| c | c | c | c |}
\hline
Parameter & Milky Way &  LMC & Reference \\
\hline
$M_{BULGE}$ & $1.5\times10^{10}$ M$_{\odot}$ & 0 M$_{\odot}$ & 1 ,3 \\
$M_{DISK}$ & $5.0\times10^{10}$ M$_{\odot}$ & $3.2 \times 10^9$ M$_{\odot}$ & 1, 4 \\
$a$ & 4.0 kpc & 1.5 kpc & 2, 5 \\
$b$ & 0.5 kpc & 0.67 kpc & 2, 5 \\
$R_{CIRC}$ & 8.5 kpc & 8.9 kpc & 1, 4 \\
$V_{CIRC}$ & 220 km s$^{-1}$ & 65 km s$^{-1}$ & 1, 4 \\
$r_o$ & 8.8 kpc & 6.3 kpc & 6, 6 \\
\hline
\end{tabular}
\end{table}

We model two different galaxies here: the Milky Way and a dwarf galaxy. The parameters for each galaxy are listed in Table \ref{tab:parameters}. To choose appropriate parameters for the dwarf galaxy we try to emulate the potential of the Large Magellanic Cloud (LMC). The parameters for the LMC's disk are often quoted for a model different from the Miyamoto disk. This alternative model for the distribution of a galaxy disk is given by \citet{b25} in terms of the luminosity instead of mass:
\begin{equation}
L(R,z)= L_oe^{-R/h}sech^2(z/z_o)
\end{equation}
where $L_o$ is the luminosity at the centre of the disk, h is the scale length of the disk and $z_o$ is the scale height of the disk. The scale length and height have previously been estimated for the LMC as $h = 1.4$ kpc \citep{b26} and $z_o = 0.5$ kpc \citep{b7}. The shape of the light distribution given by these parameters was compared to the shape of the mass distribution of the Miyamoto \& Nagai disk for different values of the $a$ and $b$ parameters to estimate what the correct $a$ and $b$ are for the LMC.\\

The code keeps track of all stars within two $RTIDE$ of the cluster's centre. In all of our simulations, the value of $RTIDE$ is initially set to the Jacobi radius of the cluster given by Equation \ref{eq:rj}. When the cluster's mass changes throughout the simulation $RTIDE$ is adjusted accordingly. When switching the cluster from the dwarf galaxy to the Milky Way we calculate the Jacobi radius given by each of the potentials and set $RTIDE$ to the larger of these two values. This method ensures that stars bound to the cluster are not accidentally removed from the simulation. To impose the tidal field of the galaxy onto the cluster, $RTIDE$ is usually taken as the radial cut off for the cluster. However, since we would like to monitor how the tidal field of the galaxy is affecting the size of the cluster as it moves from the dwarf galaxy to the Milky Way we employ a different criterion for cluster membership. Once the simulation is complete, we calculate whether or not each star is bound to the cluster. This is done by calculating kinetic energy, in the frame of the cluster, the potential energy due to all the other stars, $\Phi_c$, as well as the tidal energy term, $\Phi_T$. $\Phi_T$ accounts for both the gravitational potential of the galaxy as well as the centrifugal force that arises from using the non-inertial frame of the cluster as the reference frame \citep{b3}. A star is considered bound to the cluster if the following criterion is met:
\begin{equation}
 \frac{1}{2}\dot{x}^2 + \frac{1}{2}\dot{y}^2 + \frac{1}{2}\dot{z}^2 + \Phi_c - \Phi_T < 0
\end{equation}
Once cluster membership is determined the half mass radius is calculated from the bound stars. When the cluster is switched from the dwarf to the Milky Way, the cluster membership may change; some stars that were previously bound become unbound or vice versa. These stars are still a part of the simulation and their gravitational effects are taken into consideration as long as they remain within two $RTIDE$ of the cluster's centre. Simulations are terminated when less than 100 stars remain in the simulation or when 14 Gyr has been reached. \\   

As a basis for other simulations, an initial simulation is performed with the cluster evolving in the dwarf galaxy for 14 Gyr. The cluster is put on a circular orbit in the plane of the disk, 4 kpc from the centre of the galaxy. Once this initial simulation is complete the key characteristics of the cluster, primarily the positions and velocities of the stars, at different points in its lifetime can be extracted. We then input the cluster into a new simulation where it is given a circular orbit in the plane of the disk at different positions in the Milky Way galaxy and evolves for the rest of its lifetime in the Milky Way. We use this instantaneous change in galactic potential as an approximation of the potential a cluster would feel if it was brought into the Milky Way by a merger with a dwarf galaxy. This scenario represents the evolution of the cluster before and after the merger occurs and does not account for the continuously changing potential experienced by the cluster during the merging of the galaxies. However, these models represent the first step to understanding the full evolution of a cluster in a galaxy merger. We make the ``switch'' at a number of different points in the cluster's lifetime: at 3, 4.5 and 6 Gyr respectively and also for four different distances from the centre of the Milky Way: 6, 10, 30 and 100 kpc. We perform some additional simulations to test the robustness of our results. An identical suite of simulations is done for a cluster with the same parameters as described above but with 10,000 stars. We also perform a simulation of a cluster moved from the dwarf to the Milky Way at 6 kpc after 9 Gyr. Finally, we try giving the cluster a different orbit in the dwarf galaxy before moving it to the Milky Way at 30 kpc.

\section{Results}

\begin{figure}
\begin{center}
\includegraphics[width = \columnwidth]{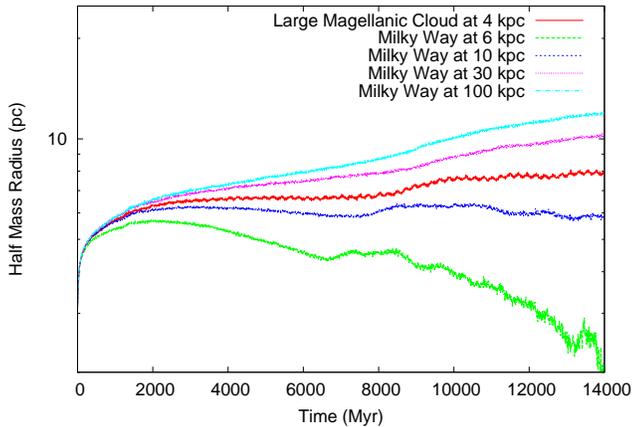}
\end{center}
\caption{Half mass radius of the cluster over time in the LMC and in the Milky Way on four different orbits. Red represents evolution in the LMC on a circular orbit at 4 kpc. Green, blue, purple and cyan represent evolution in the Milky Way on a circular orbit of 6, 10, 30 and 100 kpc respectively.}
\label{fig:ALL}
\end{figure}

To show how different galactic potentials affect the cluster's evolution, Figure \ref{fig:ALL} displays the evolution of the half mass radius over time when the cluster evolves in the LMC potential and in the Milky Way at the four galactocentric distances. The cluster experiences an initial expansion in each simulation as the stars lose mass due to stellar evolution and two-body relaxation occurs. After this initial expansion, we see that the cluster has a distinct size depending on which potential it is in. If the cluster evolves on a circular orbit at 6 or 10 kpc in the Milky Way, stars are stripped away from cluster quickly and the half mass radius decreases after the initial expansion. In contrast, on an orbit at 30 or 100 kpc the cluster continues to expand, losing fewer stars to tidal stripping because the potential is shallower at these points in the galaxy and stars farther away from the centre of the cluster can remain bound. The tidal forces experienced by the cluster in the LMC at 4 kpc are roughly equal to those at 15 kpc in the Milky Way. In all cases, the cluster experiences core collapse between 7 and 10 Gyr. This event always causes an increase in half mass radius due to the heating of the cluster from binary stars in the core \citep{b8}.\\

\begin{figure}
\begin{center}
\includegraphics[width = \columnwidth]{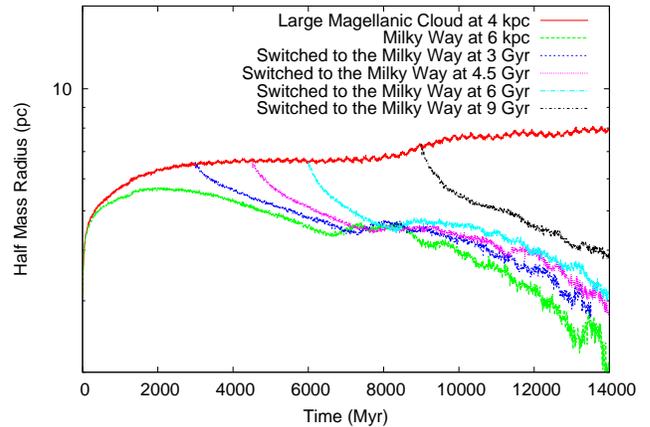}
\end{center}
\caption{Half mass radius over time for the cluster evolving entirely in the LMC at 4 kpc, the Milky Way at 6 kpc and when switched from the LMC to the Milky Way at 6 kpc. The red line represents evolution entirely in the dwarf galaxy. The green line represents the simulation of the cluster when it evolves entirely in the Milky Way, orbiting at a distance of 6 kpc from the centre. The dark blue, purple, cyan and black lines represent the simulations where the cluster is taken from the dwarf galaxy at 3, 4.5, 6 and 9 Gyr respectively and put into the Milky Way at a distance of 6 kpc from the centre.}
\label{fig:6kpc}
\end{figure}

\begin{figure}
\begin{center}
\includegraphics[width = \columnwidth]{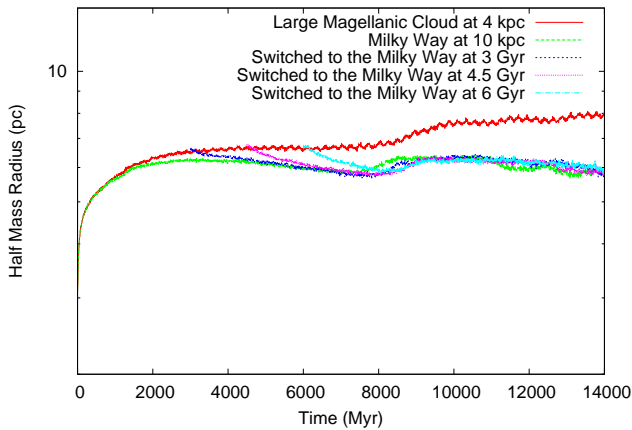}
\end{center}
\caption{Colours represent the same simulations as Figure \ref{fig:6kpc} but with all orbits in the Milky Way having a radius of 10 kpc.}
\label{fig:10kpc}
\end{figure}

Figure \ref{fig:6kpc} shows the half mass radius evolution in the LMC, the Milky Way at 6 kpc and for four simulations where the cluster is taken from the LMC and put in the Milky Way at 6 kpc at different times. When the cluster is moved to the Milky Way it adjusts quickly to the new potential with a sharp decrease in half mass radius. This quick reaction is due to the tidal field being stronger in the Milky Way at this position than in the LMC. The limiting radius of the cluster is larger than the new Jacobi radius and stars previously bound to the cluster can be stripped away. After a brief period, the accreted cluster's half mass radius starts to evolve in parallel to the half mass radius of the cluster that has lived its entire life in the Milky Way. The nature of this change in size is similar regardless of when the cluster is taken from the LMC but the timescales over which these adjustments occur are slightly different. When the cluster is switched at a later time it takes longer to adjust and ultimately the size difference between the accreted cluster and the cluster that always lives in the Milky Way is a bit larger. This result can be attributed to differences in cluster mass. When the cluster is taken from the LMC at a later time the difference between its current mass and the mass it would have if it was living in the Milky Way is larger. Adjustments take longer to occur because a greater amount of mass is stripped. However, within 1-2 half mass relaxation times of the cluster in the LMC, the size of the cluster is such that any remaining differences would not be distinguishable in observations. The same effect can be seen in Figure \ref{fig:10kpc} which shows the same set of simulations but instead the cluster is placed on an orbit at 10 kpc in the Milky Way. In this case, the cluster takes less time to adjust because the difference in the potentials is smaller and after a short time the size of the cluster is essentially what it would have been if it had always lived in the Milky Way. \\

\begin{figure}
\begin{center}
\includegraphics[width = \columnwidth]{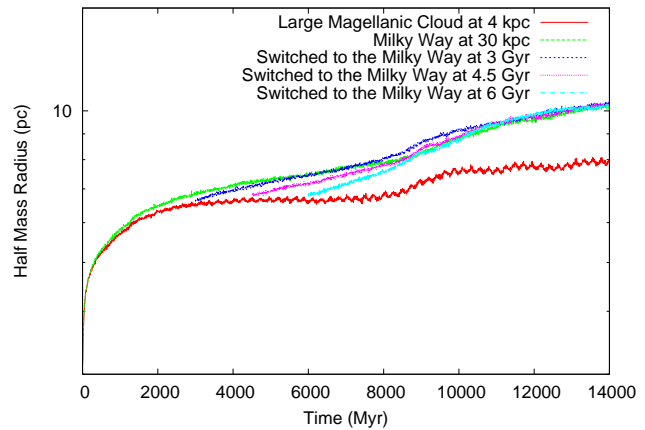}
\end{center}
\caption{Same as Figure \ref{fig:10kpc} but with all orbits in the Milky Way having a radius of 30 kpc.}
\label{fig:30kpc}
\end{figure}

\begin{figure}
\begin{center}
\includegraphics[width = \columnwidth]{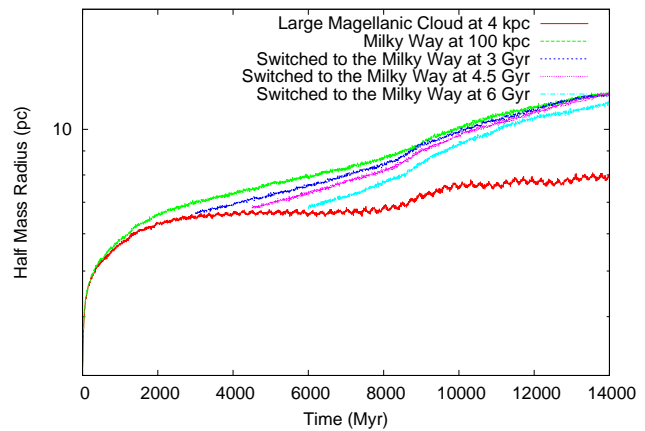}
\end{center}
\caption{Same as Figure \ref{fig:10kpc} but with all orbits in the Milky Way having a radius of 100 kpc.}
\label{fig:100kpc}
\end{figure}

When the cluster is moved from the LMC to an orbit with a large galactocentric radius in the Milky Way it experiences a reduction in tidal forces. Figures \ref{fig:30kpc} and \ref{fig:100kpc} illustrate the effect this has on cluster size. Figure \ref{fig:30kpc} shows the half mass radius of the cluster if it moves from the LMC to a 30 kpc orbit in the Milky Way. The cluster's half mass radius increases rapidly to match the half mass radius of a cluster evolving solely in the Milky Way at 30 kpc. The timescale over which this change occurs is similar to the timescale for adjustments at smaller galactocentric radii, approximately 1-2 half mass relaxation times. In this case, when the cluster is switched the limiting radius of the cluster is much smaller than the new Jacobi radius. The tidal forces are no longer sufficient to strip stars from the cluster and as stars move outwards they remain bound at larger distances. Subsequently the evolution of the half mass radius is the same as that of the cluster born in the Milky Way. A similar effect is seen when the cluster is switched and put on an orbit of 100 kpc. Adjustments take slightly longer on this orbit compared to the orbit at 30 kpc since the difference in tidal forces between the Milky Way and the LMC are larger on this orbit. We also see that increasing the time the cluster is taken from the LMC leads to a slower adjustment, similar to what is exhibited in Figure \ref{fig:6kpc}. In particular, the cluster that is taken from the LMC at 6 Gyr shows a difference of $\sim0.5$ pc after 14 Gyr while the clusters taken at earlier times overlap completely with evolution in the Milky Way. This variation can again be contributed to differences in cluster mass at these times. However, we note again that these size differences would go unnoticed in observations. \\

Similar results are obtained when we perform the same simulations with a cluster of 10,000 stars instead of 50,000. Since the same half mass radius and average mass per star are used in this simulation, we obtain a cluster with a lower mass as well as a lower mass density. When the cluster is taken from the dwarf galaxy and placed on an orbit in the Milky Way it experiences a dramatic change in half mass radius and then begins to evolve as it would had it always lived in the Milky Way. This effect is comparable to what we find with the 50,000 star cluster, the main difference being that the changes in size tend to occur over shorter timescales for the 10,000 star cluster.\\

To probe the effect of moving the cluster at later times in its dynamical evolution, we perform a simulation of a cluster that is moved from the dwarf galaxy at 9 Gyr, after core collapse has occured. The cluster is put on an orbit 6 kpc from the centre of the Milky Way. The results are represented by the black line of Figure \ref{fig:6kpc}. The cluster undergoes a large decrease in size, as it does when it is moved at earlier times. When 14 Gyr is reached, the cluster is larger, by at most 2 pc, than any of the clusters switched at an earlier time and the cluster that has always lived in the Milky Way. However, the cluster is closer in size to the Milky Way cluster and changes to its half mass radius are occurring at the same rate as the Milky Way cluster. These discrepancies in size can again be related to the mass difference of the LMC cluster and the Milky Way cluster at the time that the LMC cluster is switched to the Milky Way. By 9 Gyr, the clusters in the two galaxies have a larger difference in mass than at previous times and the cluster from the dwarf galaxy retains some of this extra mass when it is switched to the Milky Way.\\

We have also investigated the effect of putting the cluster at several different galactocentric distances in the LMC before moving it to the Milky Way. We find the adjustment of the cluster's size to the new Milky Way potential occurs regardless of where it was orbitting in the LMC.\\

\begin{figure}
\begin{center}
\includegraphics[width = \columnwidth]{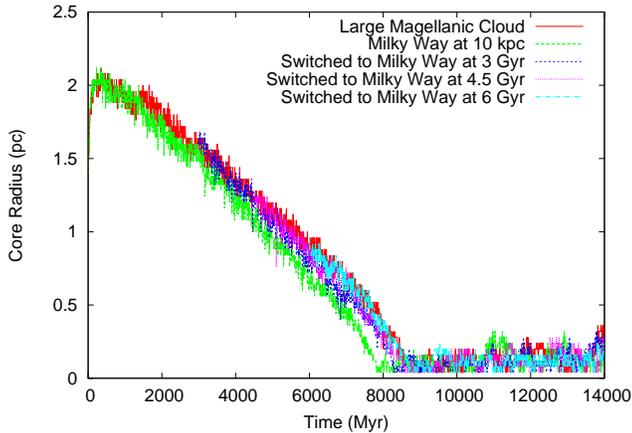}
\end{center}
\caption{Core radius over time with all orbits in the Milky Way at 10 kpc. Colours are representative of the same simulations as in Figure \ref{fig:10kpc}.}
\label{fig:Core}
\end{figure}

To examine how changes in potential can affect the inner region of a cluster, Figure \ref{fig:Core} displays the core radius over time for the set of simulations depicted in Figure \ref{fig:10kpc}, where the cluster is put on an orbit of 10 kpc in Milky Way. We see that there exists little difference between the core radius of the cluster in the LMC and core radius of the cluster that evolves entirely in the Milky Way. When the cluster is switched from the LMC to the Milky Way the core radius does not change. Such small differences in core radius are an indication that the cluster's inner regions are not sensitive to the tidal conditions that the cluster lives in. The simulations carried out at different galactocentric distances in the Milky Way give the same results.

\section{Summary and Discussion}

We present $N$-body simulations of a cluster which forms in an LMC-like dwarf galaxy and then is moved into the Milky Way after 3, 4.5 and 6 Gyr. The cluster is placed on a circular orbit in the Milky Way at four different galactocentric distances, 6, 10, 30 and 100 kpc. We find that when a dwarf galaxy cluster is moved to the Milky Way, the cluster exhibits a quick reaction to this change in environment. If the cluster experiences an increase in tidal forces, such as those felt at 6 or 10 kpc in the Milky Way, the half mass radius of the cluster will decrease rapidly as mass is lost through tidal stripping. If the cluster experiences a decrease in tidal forces, such as those at 30 or 100 kpc, the half mass radius of the cluster will increase as the mass loss rate slows due to the galaxy's inability to strip stars closer to the cluster centre. For every galactocentric distance, the cluster will ultimately reach a stage where its size appears as it would if the cluster was always resident to the Milky Way. We have also demonstrated that the effect is qualitatively independent of the time the cluster is transferred to the Milky Way. This result indicates that the stage of dynamical evolution of the cluster does not alter the influence of the Milky Way on the cluster's dynamics.\\

Based on our results, if the Milky Way were to gain several clusters in a merger event with a dwarf galaxy, after a brief period of time the distribution in half mass radii of the clusters may be indistinguishable from a size distribution of clusters that have always orbited in the Milky Way. Other characteristics such as HB morphology, age and metallicity, would have to be used to distinguish them as clusters that truly came from a dwarf galaxy. However, an observation of a cluster that appeared to have a half mass radius too small or too large for its current external potential could be an indication of a recent merger. In this case, the timescales over which size adjustments occur could potentially be used to constrain the time elapsed since a merger event.\\ 

It is interesting to note that if a cluster is accreted from a dwarf galaxy with a sufficiently strong tidal field, such as the one we have simulated here, and takes up an orbit with a large galactocentric radius, its half mass radius will actually increase. This is important when considering clusters in the Milky Way with large galactocentric distances and large half mass radii such as AM 1, Pal 3, Pal  4, Pal 14 and NGC 2419 \citep{b9}). We have shown that a large half mass radius is not necessarily an indication that the cluster came from a dwarf galaxy. The large half mass radii of these clusters can be attributed to the weak tidal field in which they currently live and large initial half mass radii. This result is consistent with the conclusions of \citet{b33} who performed $N$-body simulations of Palomar 4 with varying initial parameters to reproduce currently observed values. They estimated the initial half mass radius was roughly 10 pc and suggested this could be due to the cluster being born in a weak tidal field. \\

Our results are particularly interesting when compared with the findings of \citet{b2}. They separated the Milky Way clusters with galactocentric radii $>$ 8 kpc into two groups: one tidally underfilling group with $\frac{r_h}{r_J} < 0.05$ and one tidally filling group with $0.1 < \frac{r_h}{r_J} < 0.3$. They found no correlation between membership in the Young and Old Halo populations and membership in the tidal filling/underfilling populations. This lack of correlation indicates that accreted clusters do not preferentially belong to either the tidally filling or underfilling group. These results are consistent with our findings which show that accreted clusters should have the same distribution of  $\frac{r_h}{r_J}$ as clusters born in the Milky Way if the distribution of initial cluster masses, densities, etc. are similar. \\

We have also found, in contrast to the half mass radius, that the core radius is not greatly affected by changes in tidal forces on the cluster. This is consistent with the simulations performed by \citet{b28} which showed that the slope of the mass function in the inner regions of the cluster is roughly independent of the orbit the cluster has in the Milky Way. This result is an indication that the core radius of a cluster will depend on the initial structural conditions of the cluster and will not be affected by its tidal history. Therefore, if initial conditions in dwarf galaxies are distinct from those in the Milky Way, the core radii of accreted clusters should be different from Milky Way clusters. This is consistent with the aforementioned observational findings of \citet{b15} that demonstrated current dwarf galaxy clusters have a similar distribution of core radii to the Young Halo population in the Milky Way.\\

We simulate a cluster with 50,000 stars, the size of a small globular cluster. However, we expect that the results are qualitatively independent of the number of stars in the cluster and should hold for larger systems. We also do not expect different cluster orbits to appreciably change the results we have obtained by using a circular orbit in the plane of the disk. We have shown that only the current tidal conditions of the cluster are important to its size. Hence, even clusters accreted on more realistic orbits, such as eccentric orbits, like those studied by \citet{b28}, and inclined orbits (Webb et al., 2014b, submitted) will be similar to their Milky Way counterparts. \\

In these simulations we use a single abrupt change to the galactic potential to represent a merger between the Milky Way and a dwarf galaxy. Using this model for the potential, we have probed how the tidal field of the Milky Way effects the cluster's evolution after the merger event is complete. We have not studied the effect of the continuously changing tidal field that the cluster would experience during the merging of the two galaxies. In particular, it is unlikely that a cluster would be as close to the centre of the Milky Way as 6 kpc during or immediately after the merger. The cluster must pass through larger galactocentric distances before settling on an orbit close to the centre of the galaxy. This continuous change in potential is expected to have some effect on the cluster. Our future work will include modelling the complete time dependent potential of the merging galaxies in order to account for this evolution.

\section*{Acknowledgments}
We acknowledge the use of Sharcnet resources in the completion of this work. We would like to thank Anna Sippel for her helpful suggestions. We would also like thank the referee for their constructive comments. A.S. and J.W. are supported by NSERC.

\label{lastpage}

\end{document}